\documentclass[aps,floatfix,superscriptaddress,preprintnumbers,showpacs,showkeys]{revtex4}
\usepackage{amssymb}
\usepackage{amsmath}
\usepackage{amsfonts}
\usepackage{epsfig}
\usepackage{graphicx}
\usepackage{tabularx}
\usepackage{verbatim}
\newcommand{\eq}{\begin{eqnarray}}
\newcommand{\en}{\end{eqnarray}}
\begin{document}
\title{Effect of the $\eta\eta$ channel and interference phenomena in the two-pion
transitions of charmonia and bottomonia}

\author{Yury S. Surovtsev}
\affiliation{Bogoliubov Laboratory of Theoretical Physics,
Joint Institute for Nuclear Research, 141980 Dubna, Russia}
\author{P. Byd\v{z}ovsk\'y}
\affiliation{Nuclear Physics Institute of the CAS, 25068 \v{R}e\v{z},
Czech Republic}
\author{Thomas Gutsche}
\affiliation{Institut f\"ur Theoretische Physik,
Universit\"at T\"ubingen,
Kepler Center for Astro and Particle Physics,
Auf der Morgenstelle 14, D-72076 T\"ubingen, Germany}
\author{Robert Kami\'nski}
\affiliation{Institute of Nuclear Physics of the PAN, Cracow 31342, Poland}
\author{Valery E. Lyubovitskij}
\affiliation{Institut f\"ur Theoretische Physik,
Universit\"at T\"ubingen,
Kepler Center for Astro and Particle Physics,
Auf der Morgenstelle 14, D-72076 T\"ubingen, Germany}
\affiliation{Department of Physics, Tomsk State University,
634050 Tomsk, Russia}
\affiliation{Mathematical Physics Department,
Tomsk Polytechnic University,
Lenin Avenue 30, 634050 Tomsk, Russia}
\author{Miroslav Nagy}
\affiliation{Institute of Physics, SAS, Bratislava 84511, Slovak Republic}

\date{\today}

\begin{abstract}
The basic shape of di-pion mass spectra in the two-pion transitions of both charmonia and bottomonia
states is explained by an unified mechanism based on contributions of the $\pi\pi$, $K\overline{K}$
and $\eta\eta$ coupled channels including their interference. The role of the individual $f_0$ resonances
in shaping the di-pion mass distributions in the charmonia and bottomonia decays is considered.
\end{abstract}

\pacs{11.55.Bq,11.80.Gw,12.39.Mk,14.40.Pq}

\keywords{coupled--channel formalism, meson--meson scattering,
heavy meson decays, scalar and pseudoscalar mesons}

\maketitle
\section{Introduction}

In this contribution we report on new results in continuation of our study of
scalar meson properties analyzing jointly the data on the isoscalar S-wave processes
$\pi\pi\to\pi\pi,K\overline{K},\eta\eta$ and on the two-pion transitions of heavy mesons,
when it is reasonable to consider that the two-pion pair is produced in the S-wave
state and the final meson remains a spectator.
We analyzed data on the charmonium decay processes --- $J/\psi\to\phi\pi\pi$, $\psi(2S)\to J/\psi\pi\pi$ ---
from the Crystal Ball, DM2, Mark~II, Mark~III, and BES~II Collaborations and practically all
available data on two-pion transitions of the $\Upsilon$ mesons from the ARGUS, CLEO, CUSB, Crystal Ball,
Belle, and {\it BaBar} Collaborations --- $\Upsilon(mS)\to\Upsilon(nS)\pi\pi$ ($m>n$, $m=2,3,4,5,$ $n=1,2,3$).
Moreover, the contribution of multi-channel $\pi\pi$ scattering (namely,
$\pi\pi\to\pi\pi,K\overline{K},\eta\eta$) in the final-state interactions is considered.
The multi-channel $\pi\pi$ scattering is described in our model-independent approach based on analyticity
and unitarity and using an uniformization procedure.
A novel feature in the analysis is accounting for effects from the $\eta\eta$ channel
in the indicated two-pion transitions which is assumed not only kinematically, i.e., via including
the channel threshold in the uniformizing variable, but also by adding the $\pi\pi\to\eta\eta$
amplitude in the corresponding formulas for the decays.

We showed that the experimentally observed interesting (even mysterious) behavior of the $\pi\pi$ spectra
of the $\Upsilon$-family decays, which starts to be apparent in the second radial excitation and
is also seen for the higher states, --- a bell-shaped form in the near-$\pi\pi$-threshold region,
smooth dips at about 0.6~GeV in the $\Upsilon(4S,5S)\to\Upsilon(1S) \pi^+ \pi^-$, about 0.45~GeV in
the $\Upsilon(4S,5S)\to\Upsilon(2S) \pi^+ \pi^-$, and at about 0.7~GeV in the
$\Upsilon(3S)\to\Upsilon(1S)(\pi^+\pi^-,\pi^0\pi^0)$, and also sharp dips near 1~GeV
in the $\Upsilon(4S,5S)\to\Upsilon(1S) \pi^+ \pi^-$ ---
can be explained by the interference between the $\pi\pi$ scattering amplitudes,
$K\overline{K}\to\pi\pi$ and $\eta\eta\to\pi\pi$, in the final-state re-scattering
(by the constructive interference in the near-$\pi\pi$-threshold region and by the destructive one
in the dip regions).
Note that in a number of works (see, e.g., \cite{Simonov-Veselov} and the references therein) various
assumptions were made to explain this observed behavior of the di-meson mass distributions.

We have explained the basic shape of di-pion mass spectra in the two-pion transitions of both charmonia and bottomonia on the basis of our previous conclusions on wide resonances without any additional assumptions.
In works \cite{SBLKN-jpgnpp14,SBLKN-prd14} we have shown:
If a wide resonance cannot decay into a channel which opens above its mass, but the resonance is strongly coupled to this channel, then one should consider this resonance as a multi-channel state allowing for this closed channel.

\section{The effect of multi-channel $\pi\pi$ scattering in decays of the $\psi$- and $\Upsilon$-meson families}

We have considered the three-channel case of the multi-channel $\pi\pi$ scattering, i.e. the reactions
$\pi\pi\to\pi\pi,K\overline{K},\eta\eta$, because it was shown \cite{SBLKN-jpgnpp14} that this
is a minimal number of coupled channels needed for obtaining correct values of $f_0$-resonance parameters.
In the combined analysis, the data for the multi-channel $\pi\pi$ scattering were taken from
many papers (see Refs. in~\cite{SBLKN-prd14}). For the decay $J/\psi\to\phi\pi^+\pi^-$ data
were taken from Mark III, DM2 and BES II Collaborations;
for $\psi(2S)\to J/\psi(\pi^+\pi^-~{\rm and}~\pi^0\pi^0)$ --- from Mark~II and Crystal Ball(80)
(see Refs. also in~\cite{SBLKN-prd14}).
For $\Upsilon(2S)\to\Upsilon(1S)(\pi^+\pi^-~{\rm and}~\pi^0\pi^0)$ data were used from ARGUS~\cite{Argus}, CLEO~\cite{CLEO}, CUSB~\cite{CUSB}, and Crystal Ball~\cite{Crystal_Ball(85)} Collaborations; for
$\Upsilon(3S)\to\Upsilon(1S)(\pi^+\pi^-,\pi^0\pi^0)$ and $\Upsilon(3S)\to\Upsilon(2S)(\pi^+\pi^-,\pi^0\pi^0)$ --- from CLEO \cite{CLEO(94),CLEO07}; for $\Upsilon(4S)\to\Upsilon(1S,2S)\pi^+\pi^-$ --- from {\it BaBar} \cite{BaBar06} and Belle \cite{Belle}; for $\Upsilon(5S)\to\Upsilon(1S,2S,3S)\pi^+\pi^-$ --- from Belle Collaboration~\cite{Belle}.

The di-meson mass distributions in the quarkonia decays are calculated using a formalism
analogous to that proposed in Ref.~\cite{MP-prd93} for the decays $J/\psi\to\phi(\pi\pi, K\overline{K})$
and $V^{\prime}\to V\pi\pi$ ($V=\psi,\Upsilon$) which is extended with allowing for amplitudes of
transitions between the $\pi\pi$, $K\overline{K}$ and $\eta\eta$ channels in decay formulas.
It is assumed that the pion pairs in the final state have zero isospin and spin. Only these pairs of
pions undergo the final state interactions whereas the final $\Upsilon(nS)$ meson ($n<m$) remains
as a spectator. The decay amplitudes are related with the scattering amplitudes
$T_{ij}$ $(i,j=1-\pi\pi,2-K\overline{K},3-\eta\eta)$ as follows
\begin{eqnarray}
&&F\bigl(J/\psi\to\phi\pi\pi\bigr)=c_1(s)T_{11}+\Bigl(\frac{\alpha_2}{s-\beta_2}+c_2(s)\Bigr)T_{21}+c_3(s)T_{31},\\
&&F\bigl(\psi(2S)\to\psi(1S)\pi\pi\bigr)=d_1(s)T_{11}+d_2(s)T_{21}+d_3(s)T_{31},\\
&&F\bigl(\Upsilon(mS)\to\Upsilon(nS)\pi\pi\bigr) = e_1^{(mn)}T_{11}
+ e_2^{(mn)}T_{21}+ e_3^{(mn)}T_{31},\\
&&~~~~~~~~~~~~m>n,~ m=2,3,4,5,~ n=1,2,3\nonumber
\end{eqnarray}
where $c_i=\gamma_{i0}+\gamma_{i1}s$, $d_i=\delta_{i0}+\delta_{i1}s$ and $e_i^{(mn)}=\rho_{i0}^{(mn)}+\rho_{i1}^{(mn)}s$; indices $m$ and $n$ correspond to $\Upsilon(mS)$ and $\Upsilon(nS)$, respectively. The free parameters $\alpha_2$, $\beta_2$, $\gamma_{i0}$, $\gamma_{i1}$, $\delta_{i0}$, $\delta_{i1}$, $\rho_{i0}^{(mn)}$ and $\rho_{i1}^{(mn)}$ depend on the couplings of $J/\psi$, $\psi(2S)$, and $\Upsilon(mS)$ to the channels $\pi\pi$, $K\overline{K}$ and $\eta\eta$. The pole term in eq.(1) in front of $T_{21}$ is an approximation of possible $\phi K$ states, not forbidden by OZI rules.

The amplitudes $T_{ij}$ are expressed through the $S$-matrix elements
\begin{equation}
S_{ij}=\delta_{ij}+2i\sqrt{\rho_i\rho_j}T_{ij}
\end{equation}
where $\rho_i=\sqrt{1-s_i/s}$ and $s_i$ is the reaction threshold. The $S$-matrix elements are taken as the products
\begin{equation}
S=S_B S_{res}
\end{equation}
where $S_{res}$ represents the contribution of resonances, $S_B$ is the background part.
The $S_{res}$-matrix elements are parameterized on the uniformization plane of the $\pi\pi$-scattering $S$-matrix element by poles and zeros which represent resonances. The uniformization plane is obtained by a conformal map of the 8-sheeted Riemann surface, on which the three-channel $S$ matrix is determined, onto the plane. In the uniformizing variable used~\cite{SBL-prd12}
\begin{equation}
w=\frac{\sqrt{(s-s_2)s_3} + \sqrt{(s-s_3)s_2}}{\sqrt{s(s_3-s_2)}}~~~~(s_2=4m_K^2 ~ {\rm and}~ s_3=4m_\eta^2)
\end{equation}
we have neglected the $\pi\pi$-threshold branch point and allowed for the $K\overline{K}$- and $\eta\eta$-threshold branch points and left-hand branch point at $s=0$ related to the crossed channels.

Resonance representations on the Riemann surface are obtained using formulas in Table~\ref{tab:anal.contin.} \cite{KMS-96}, expressing analytic continuations of the $S$-matrix elements to all sheets in terms of those on the physical (I) sheet that have only the resonances zeros (beyond the real axis), at least, around the physical region.
\begin{table}[!htb]
\begin{center}
\caption{Analytic continuations of the $S$-matrix elements}\label{tab:anal.contin.}
{
\large
\def\arraystretch{1.4}
\begin{tabular}{ccccccccc}\hline
{Process} & I & II & III & IV & V & VI & VII & VIII \\
\hline $1\to 1$ & $S_{11}$ & $\frac{1}{S_{11}}$ &
$\frac{S_{22}}{D_{33}}$ & $\frac{D_{33}}{S_{22}}$ & $\frac{\det
S}{D_{11}}$
& $\frac{D_{11}}{\det S}$ & $\frac{S_{33}}{D_{22}}$ & $\frac{D_{22}}{S_{33}}$\\
$1\to 2$ & $S_{12}$ & $\frac{iS_{12}}{S_{11}}$ &
$\frac{-S_{12}}{D_{33}}$ & $\frac{iS_{12}}{S_{22}}$
& $\frac{iD_{12}}{D_{11}}$ & $\frac{-D_{12}}{\det S}$ & $\frac{iD_{12}}{D_{22}}$ & $\frac{D_{12}}{S_{33}}$\\
$2\to 2$ & $S_{22}$ & $\frac{D_{33}}{S_{11}}$ &
$\frac{S_{11}}{D_{33}}$ & $\frac{1}{S_{22}}$ &
$\frac{S_{33}}{D_{11}}$
& $\frac{D_{22}}{\det S}$ & $\frac{\det S}{D_{22}}$ & $\frac{D_{11}}{S_{33}}$\\
$1\to 3$ & $S_{13}$ & $\frac{iS_{13}}{S_{11}}$ &
$\frac{-iD_{13}}{D_{33}}$ & $\frac{-D_{13}}{S_{22}}$
& $\frac{-iD_{13}}{D_{11}}$ & $\frac{D_{13}}{\det S}$ & $\frac{-S_{13}}{D_{22}}$ & $\frac{iS_{13}}{S_{33}}$\\
$2\to 3$ & $S_{23}$ & $\frac{D_{23}}{S_{11}}$ &
$\frac{iD_{23}}{D_{33}}$ & $\frac{iS_{23}}{S_{22}}$
& $\frac{-S_{23}}{D_{11}}$ & $\frac{-D_{23}}{\det S}$ & $\frac{iD_{23}}{D_{22}}$ & $\frac{iS_{23}}{S_{33}}$\\
$3\to 3$ & $S_{33}$ & $\frac{D_{22}}{S_{11}}$ & $\frac{\det
S}{D_{33}}$ & $\frac{D_{11}}{S_{22}}$ & $\frac{S_{22}}{D_{11}}$ &
$\frac{D_{33}}{\det S}$ & $\frac{S_{11}}{D_{22}}$ &
$\frac{1}{S_{33}}$\\\hline
\end{tabular}}
\end{center}
\end{table}
In Table~\ref{tab:anal.contin.} the Roman numerals denote the Riemann-surface sheets, the superscript~$I$ is omitted to simplify the notation,
$\det S$ is the determinant of the $3\times3$ $S$-matrix on sheet
I, $D_{\alpha\beta}$ is the minor of the element
$S_{\alpha\beta}$, that is, $D_{11}=S_{22}S_{33}-S_{23}^2$,
$D_{22}=S_{11}S_{33}-S_{13}^2$, $D_{33}= S_{11}S_{22}-S_{12}^2$,
$D_{12}=S_{12}S_{33}-S_{13}S_{23}$, $D_{23}=
S_{11}S_{23}-S_{12}S_{13}$, etc.

These formulas show how singularities and resonance poles and zeros are transferred from the matrix element $S_{11}$ to matrix elements of coupled processes.

The background is introduced to the $S_B$-matrix elements in a natural way: on the threshold of
each important channel there appears generally speaking a complex phase shift. It is important
that we have obtained practically zero background of the $\pi\pi$ scattering in the scalar-isoscalar
channel. First, this confirms well our assumption (5).
Second, this shows that the representation of multi-channel resonances by the pole and zeros on
the uniformization plane given in Table~1 is good and quite sufficient.
This result is also a criterion for the correctness of the approach.

In Table~\ref{tab:clusters} we show the poles corresponding to $f_0$ resonances, obtained in the analysis.
\begin{table}[!htb]
\begin{center}
\caption{The poles for resonances on the $\sqrt{s}$-plane. ~$\sqrt{s_r}\!=\!{\rm E}_r\!-\!i\Gamma_r/2$.} \label{tab:clusters}
\vspace*{-0.01cm}
{
\def\arraystretch{1.3}
\begin{tabular}{|c|c|c|c|c|c|c|c|}
\hline ${\rm Sheet}$ & {} & $f_0(500)$ & $f_0(980)$ & $f_0(1370)$ & $f_0(1500)$ & $f_0^\prime(1500)$ & $f_0(1710)$ \\ \hline
${\rm II}$ & {${\rm E}_r$} & $521.6\!\pm\!12.4$ & $1008.4\!\pm\!3.1$ & {} & {} & $1512.4\!\pm\!4.9$ & {} \\
{} & {$\Gamma_r/2$} & $467.3\!\pm\!5.9$ & $33.5\!\pm\!1.5$ & {} & {} & $287.2\!\pm\!12.9$ & {} \\
\hline ${\rm III}$ & {${\rm E}_r$} & $552.5\!\pm\!17.7$ & $976.7\!\pm\!5.8$ & $1387.2\!\pm\!24.4$ & {} & $1506.1\!\pm\!9.0$ & {} \\
{} & {$\Gamma_r/2$} & $467.3\!\pm\!5.9$ & $53.2\!\pm\!2.6$ & $167.2\!\pm\!41.8$ & {} & $127.8\!\pm\!10.6$ & {} \\
\hline ${\rm IV}$ & {${\rm E}_r$} & {} & {} & $1387.2\!\pm\!24.4$ & {} & $1512.4\!\pm\!4.9$ & {} \\
{} & {$\Gamma_r/2$} & {} & {} & $178.2\!\pm\!37.2$ & {} & $215.0\!\pm\!17.6$ & {} \\
\hline ${\rm V}$ & {${\rm E}_r$} & {} & {} & $1387.2\!\pm\!24.4$ & $1493.9\!\pm\!3.1$ & $1498.8\!\pm\!7.2$ & $1732.8\!\pm\!43.2$ \\
{} & {$\Gamma_r/2$} & {} & {} & $261.0\!\pm\!73.7$ & $72.8\!\pm\!3.9$ & $142.3\!\pm\!6.0$ & $114.8\!\pm\!61.5$ \\
\hline ${\rm VI}$ & {${\rm E}_r$} & $573.4\!\pm\!29.1$ & {} & $1387.2\!\pm\!24.4$ & $1493.9\!\pm\!5.6$
& $1511.5\!\pm\!4.3$ & $1732.8\!\pm\!43.2$ \\
{} & {$\Gamma_r/2$} & $467.3\!\pm\!5.9$ & {} & $250.0\!\pm\!83.1$ & $58.4\!\pm\!2.8$ & $179.3\!\pm\!4.0$ & $111.2\!\pm\!8.8$ \\
\hline ${\rm VII}$ & {${\rm E}_r$} & $542.5\!\pm\!25.5$ & {} & {} & $1493.9\!\pm\!5.0$ & $1500.4\!\pm\!9.3$ & $1732.8\!\pm\!43.2$ \\
{} & {$\Gamma_r/2$} & $467.3\!\pm\!5.9$ & {} & {} & $47.8\!\pm\!9.3$ & $99.9\!\pm\!18.0$ & $55.2\!\pm\!38.0$ \\
\hline ${\rm VIII}$ & {${\rm E}_r$} & {} & {} & {} & $1493.9\!\pm\!3.2$ & $1512.4\!\pm\!4.9$ & $1732.8\!\pm\!43.2$ \\
{} & {$\Gamma_r/2$} & {} & {} & {} & $62.2\!\pm\!9.2$ & $298.4\!\pm\!14.5$ & $58.8\!\pm\!16.4$ \\
\hline
\end{tabular}}
\end{center}
\end{table}
Generally, {\it the wide multi-channel states are most adequately represented by poles}, because
the poles give the main model-independent effect of resonances and are rather stable characteristics
for various models, whereas masses and total widths are very model-dependent for wide
resonances~\cite{SKN-epja02}.
The masses, widths, and the coupling constants of resonances should be calculated using the poles
on sheets II, IV and VIII, because only on these sheets the analytic continuations have the forms
(see Table~\ref{tab:anal.contin.}):
$$\propto 1/S_{11}^{\rm I},~~\propto 1/S_{22}^{\rm I}~~{\rm and}~~\propto 1/S_{33}^{\rm I},$$
respectively, i.e., the pole positions of resonances are at the same points of the complex-energy
plane, as the resonance zeros on the physical sheet, and are not shifted due to the coupling of
channels.

Further, since studying the decays of charmonia and bottomonia, we investigated the role of the
individual $f_0$ resonances in contributing to the shape of the di-pion mass distributions in these
decays, firstly we studied their role in forming the energy dependence of amplitudes of reactions
$\pi\pi\to\pi\pi,K\overline{K},\eta\eta$.
In this case we switched off only those resonances [$f_0(500)$, $f_0(1370)$,
$f_0(1500)$ and $f_0(1710)$], removal of which can be somehow compensated by
correcting the background (maybe, with elements of the pseudo-background) to have
the more-or-less acceptable description of the multi-channel $\pi\pi$ scattering.
Below we therefore considered description of the multi-channel $\pi\pi$ scattering for two more cases:
\begin{itemize}
\item
first, when leaving out a minimal set of the $f_0$ mesons consisting of the $f_0(500)$, $f_0(980)$, and $f_0^\prime(1500)$, which is sufficient to achieve a description of the processes $\pi\pi\to\pi\pi,K\overline{K},\eta\eta$ with a total $\chi^2/\mbox{ndf}\approx1.20$.
\item
Second, from the above-indicated three mesons only the $f_0(500)$ can be omitted while still
obtaining a reasonable description of multi-channel $\pi\pi$ scattering (though with
appearance of a pseudo-background) with the total $\chi^2/\mbox{ndf}\approx1.43$.
\end{itemize}

In Figure~1 we show the obtained description of the processes $\pi\pi\!\to\!\pi\pi,K\overline{K},\eta\eta$. The solid lines correspond to contribution of all relevant $f_0$-resonances; the dotted, of the $f_0(500)$, $f_0(980)$, and $f_0^\prime(1500)$; the dashed, of the $f_0(980)$ and $f_0^\prime(1500)$.
\begin{figure}[!thb]
\begin{center}
\includegraphics[width=0.495\textwidth,angle=0]{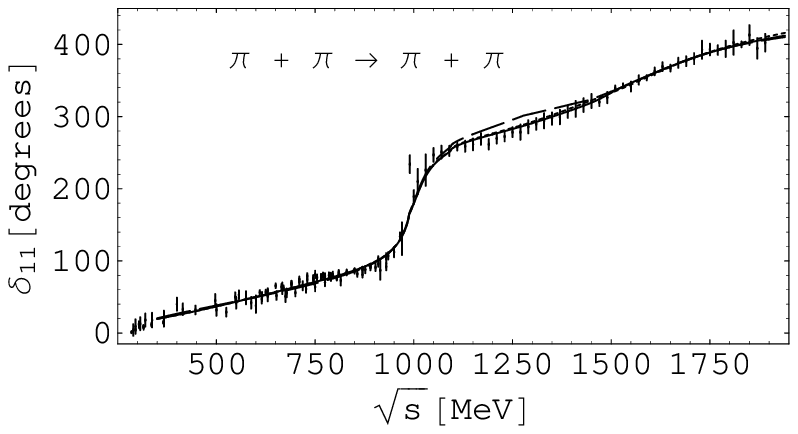}
\includegraphics[width=0.495\textwidth,angle=0]{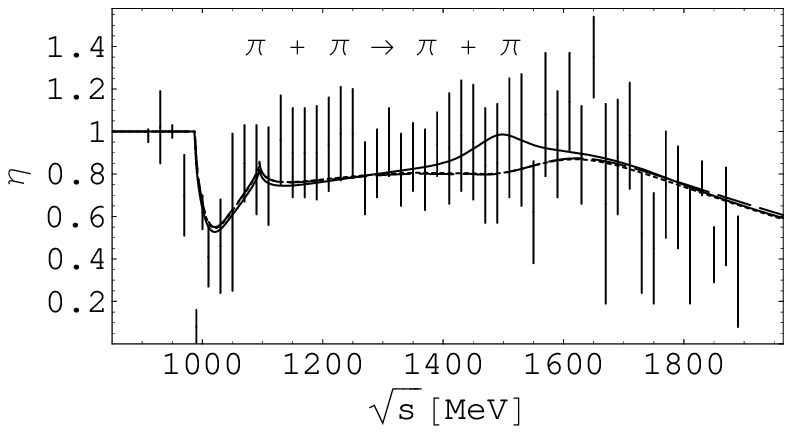}\\
\includegraphics[width=0.495\textwidth,angle=0]{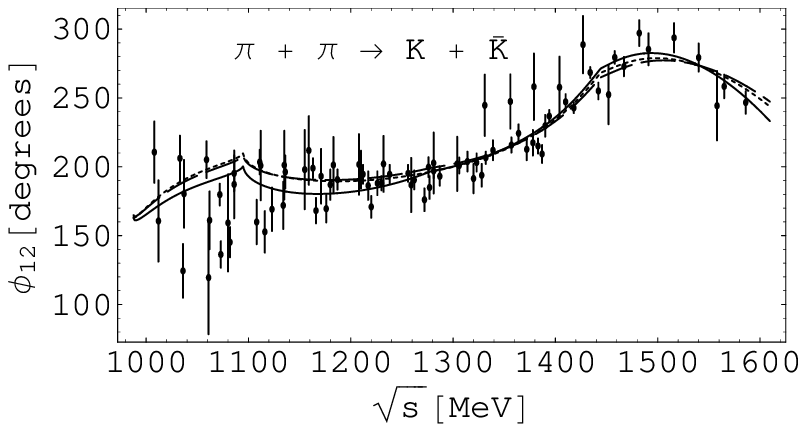}
\includegraphics[width=0.495\textwidth,angle=0]{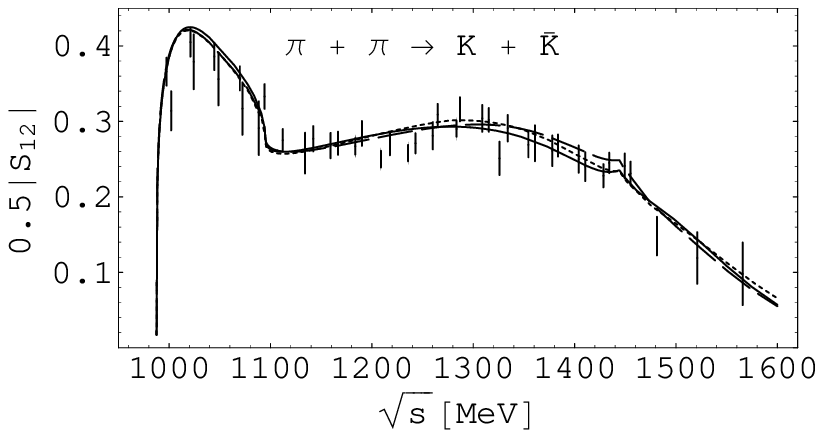}\\
\vspace*{-0.0cm}
\includegraphics[width=0.495\textwidth,angle=0]{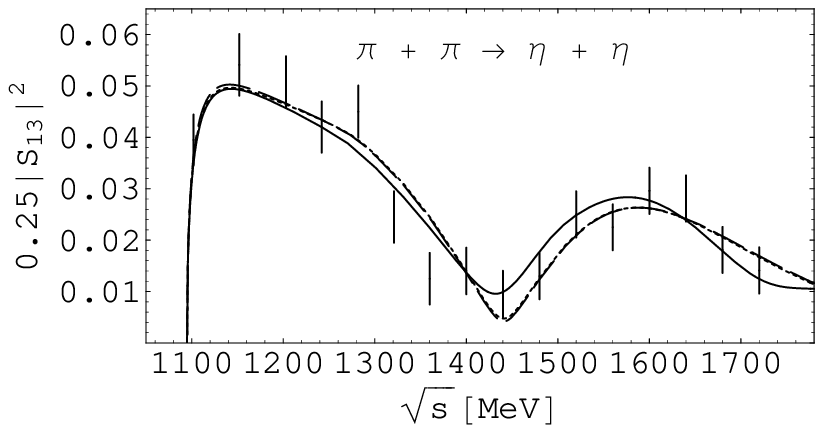}
\vskip -.3cm
\caption{The phase shifts and moduli of the $S$-matrix element in the S-wave $\pi\pi$-scattering (upper panel), in $\pi\pi\to K\overline{K}$ (middle panel), and the squared modulus of the $\pi\pi\to\eta\eta$ $S$-matrix element (lower figure).}
\end{center}\label{fig:fitting}
\end{figure}
One can see that the curves are quite similar in all three cases.

The di-meson mass distributions in the decay analysis were calculated using the relation
\begin{equation}
N|F|^{2}\sqrt{(s-s_1)[m_\psi^{2}-(\sqrt{s}-m_\phi)^{2}][m_\psi^2-
(\sqrt{s}+m_\phi)^2]}
\end{equation}
for the decay $J/\psi\to\phi\pi\pi$ and with analogous relations for $\psi(2S)\to\psi(1S)\pi\pi$
and $\Upsilon(mS)\to\Upsilon(nS)\pi\pi$.
The normalization to the experiment, $N$ is: for $J/\psi\to\phi\pi\pi$ ~0.5172 (Mark III), 0.1746 (DM 2)
and 3.8 (BES II); for $\psi(2S)\to J/\psi\pi^+\pi^-$ 1.746 (Mark II); for $\psi(2S)\to J/\psi\pi^0\pi^0$ 1.6891 (Crystal Ball(80)); for $\Upsilon(2S)\to \Upsilon(1S)\pi^+\pi^-$ 4.1758 (ARGUS), ~2.0445 (CLEO(94)) and 1.0782 (CUSB); for $\Upsilon(2S)\to\Upsilon(1S)\pi^0\pi^0$  0.0761 (Crystal Ball(85)); for $\Upsilon(3S)\to\Upsilon(1S)(\pi^+\pi^-~{\rm and}~\pi^0\pi^0)$  19.8825 and ~4.622 (CLEO(07)); for
$\Upsilon(3S)\to\Upsilon(2S)(\pi^+\pi^-$ ${\rm and}~\pi^0\pi^0)$ ~1.6987 and ~1.1803 (CLEO(94)); for $\Upsilon(4S)\to\Upsilon(1S)\pi^+\pi^-$ ~4.6827 ({\it BaBar}(06)) and ~0.3636 (Belle(07)); for $\Upsilon(4S)\to\Upsilon(2S)\pi^+\pi^-$, ~37.9877 ({\it BaBar}(06)); for $\Upsilon(5S)\to\Upsilon(1S)\pi^+\pi^-$,  $\Upsilon(5S)\to\Upsilon(2S)\pi^+\pi^-$ and $\Upsilon(5S)\to\Upsilon(3S)\pi^+\pi^-$ respectively ~0.2047, 2.8376 and 6.9251 (Belle(12)).

A satisfactory description of all considered processes (including $\pi\pi\to\pi\pi,K\overline{K},\eta\eta$)
was obtained with the total $\chi^2/\mbox{ndf}=736.457/(710 - 118)\approx1.24$;
for the $\pi\pi$ scattering, $\chi^2/\mbox{ndf}\approx1.15$.
Results for the distributions are shown in Figs. 2-4 with the same notation as in Fig.~1.
Here the effects of omitting some resonance are more apparent than in Fig.~1.
\begin{figure}[!h]
\begin{center}
\includegraphics[width=0.54\textwidth,angle=0]{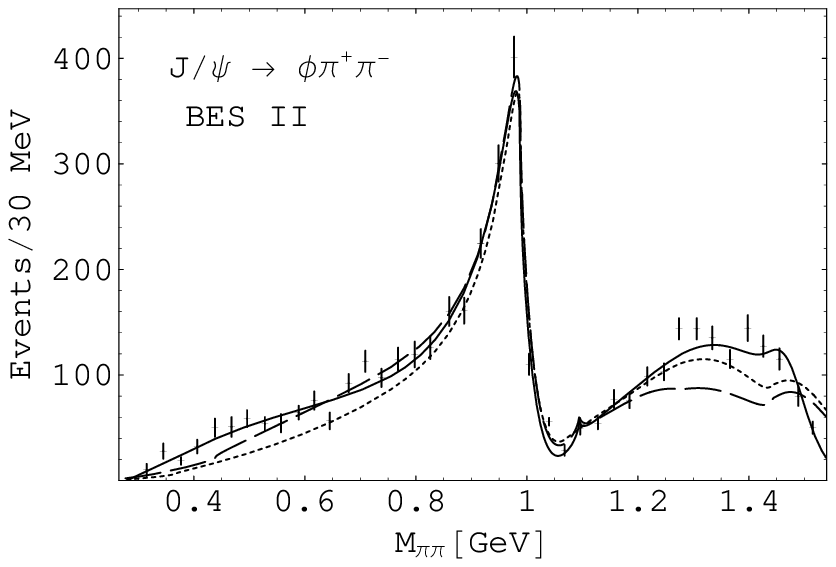}\\
\vspace*{0.3cm}
\includegraphics[width=0.48\textwidth]{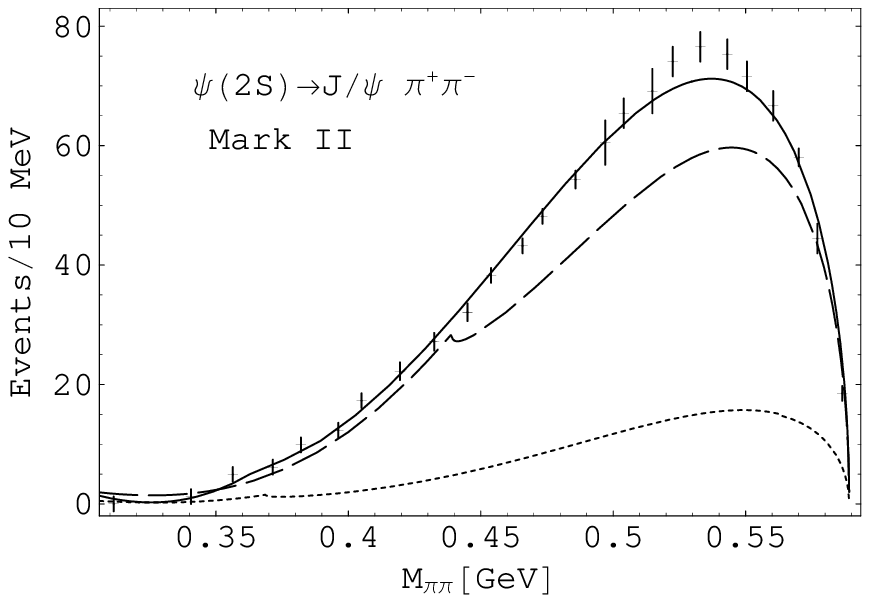}
\includegraphics[width=0.48\textwidth]{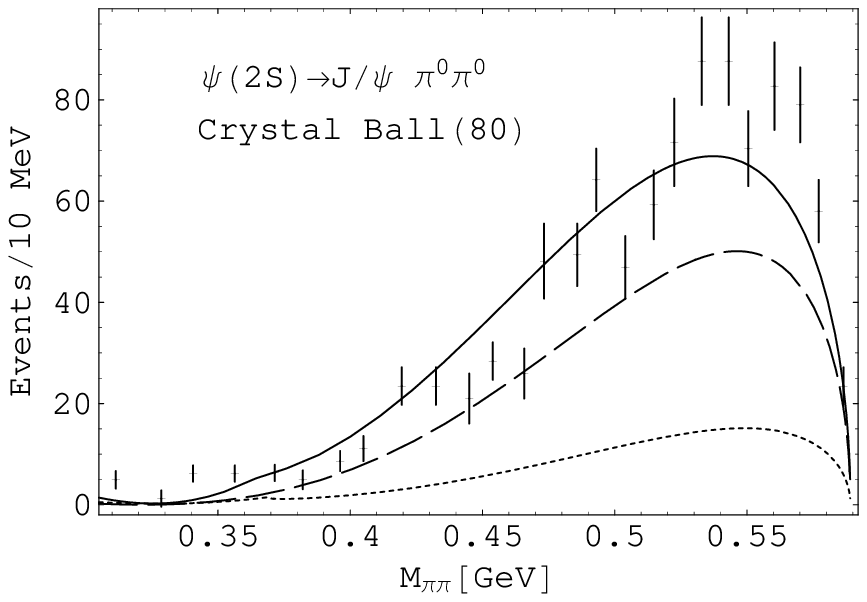}
\caption{The decays $J/\psi\to\phi\pi\pi$ and $\psi(2S)\to J/\psi\pi\pi$. The solid lines correspond to contribution of all relevant $f_0$-resonances; the dotted, of the $f_0(500)$, $f_0(980)$, and $f_0^\prime(1500)$; the dashed, of the $f_0(980)$ and $f_0^\prime(1500)$.}
\end{center}\label{fig:BESII}
\end{figure}
\begin{figure}[!h]
\begin{center}
\includegraphics[width=0.47\textwidth]{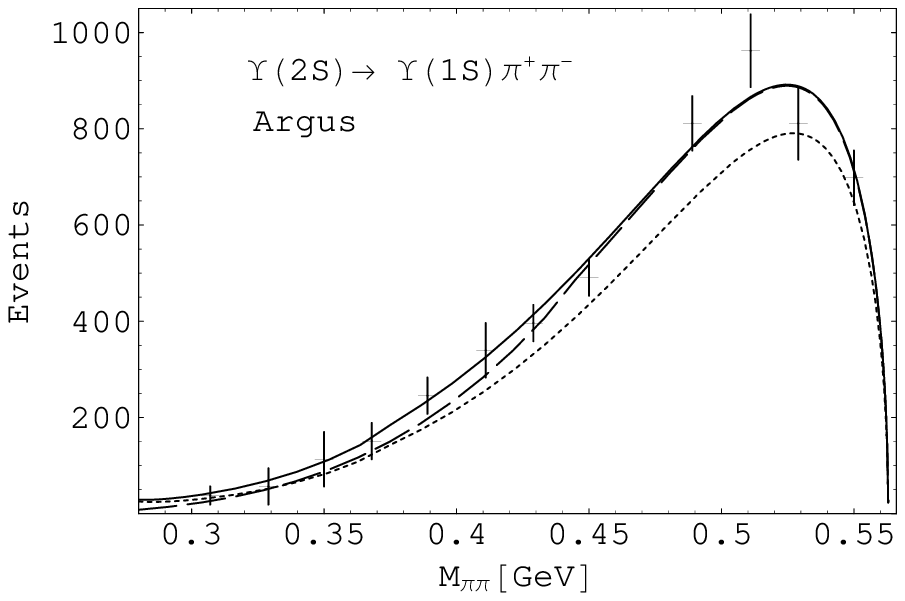}
\includegraphics[width=0.47\textwidth]{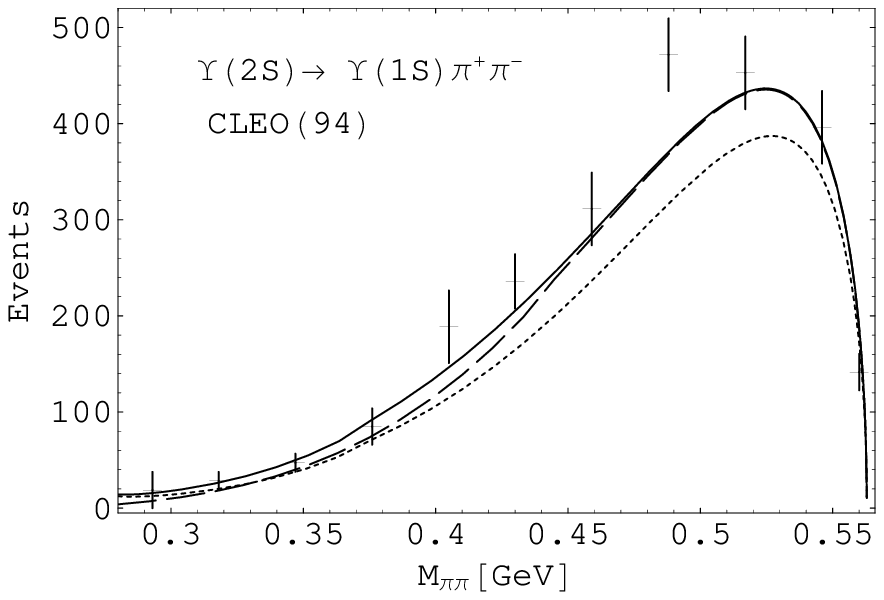}\\
\includegraphics[width=0.47\textwidth]{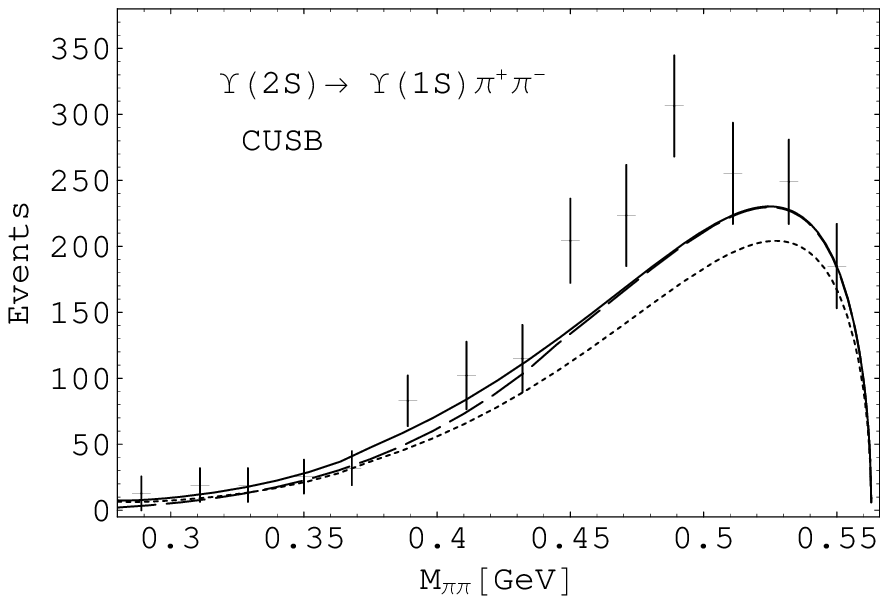}
\includegraphics[width=0.47\textwidth]{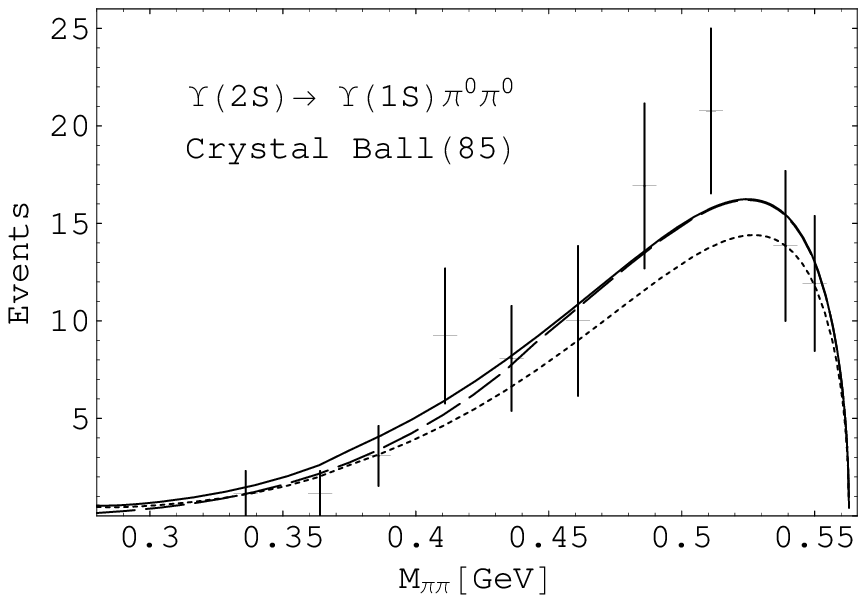}\\
\includegraphics[width=0.47\textwidth]{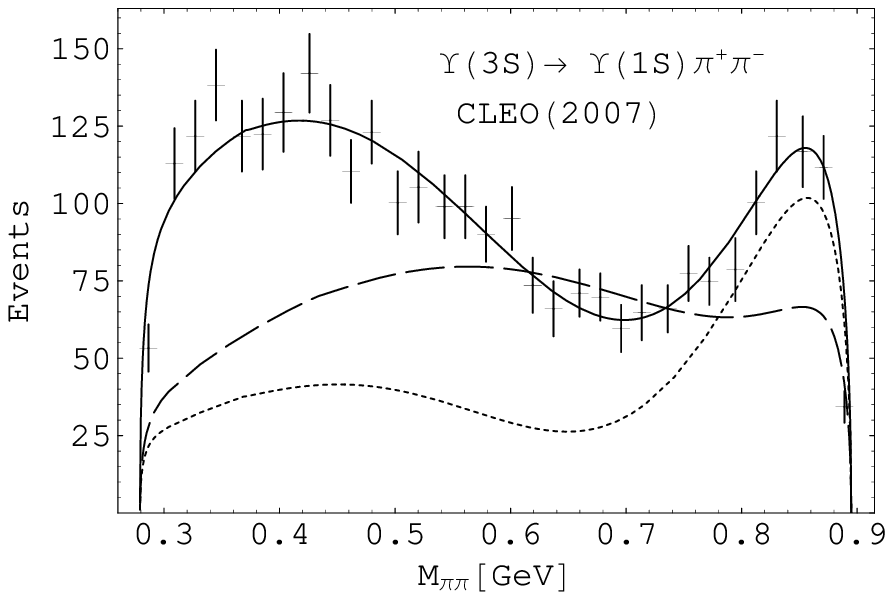}
\includegraphics[width=0.47\textwidth]{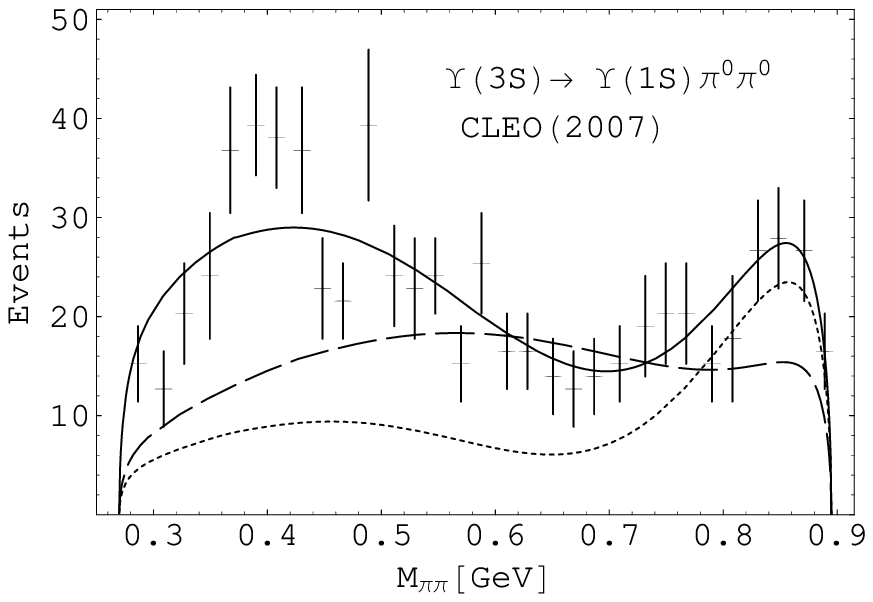}\\
\includegraphics[width=0.47\textwidth]{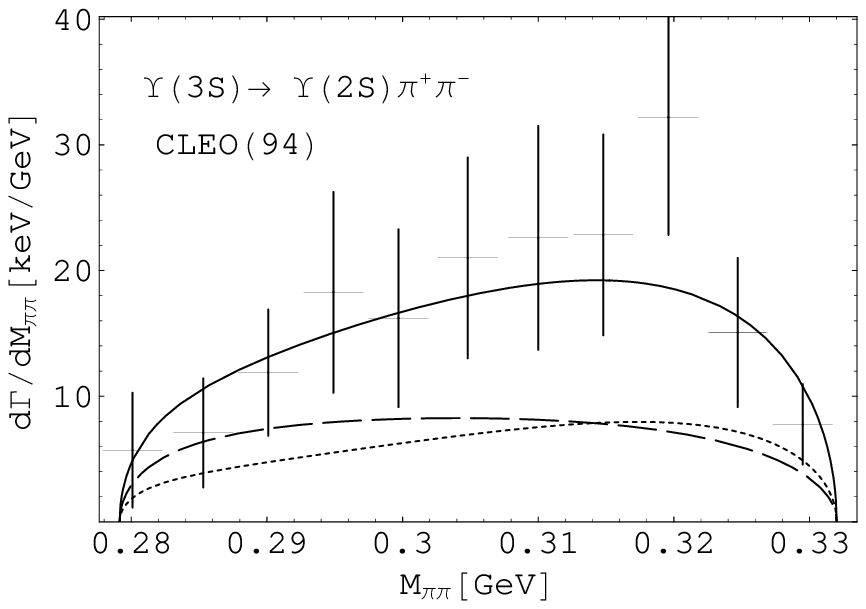}
\includegraphics[width=0.47\textwidth]{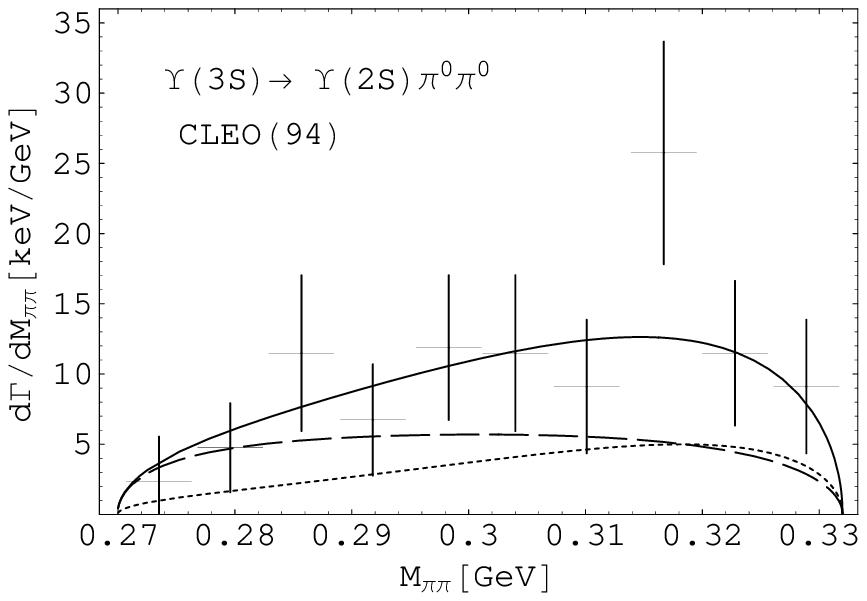}
\vspace*{-0.0cm}\caption{The decays $\Upsilon(2S)\to\Upsilon(1S)\pi\pi$ (two upper panels), $\Upsilon(3S)\to\Upsilon(1S)\pi\pi$ (middle panel) and $\Upsilon(3S)\to\Upsilon(2S)\pi\pi$ (lower panel). The solid lines correspond to contribution of all relevant $f_0$-resonances; the dotted, of the $f_0(500)$, $f_0(980)$, and $f_0^\prime(1500)$; the dashed, of the $f_0(980)$ and $f_0^\prime(1500)$.
}
\end{center}\label{fig:Ups21}
\end{figure}
\begin{figure}[!h]
\begin{center}
\includegraphics[width=0.48\textwidth]{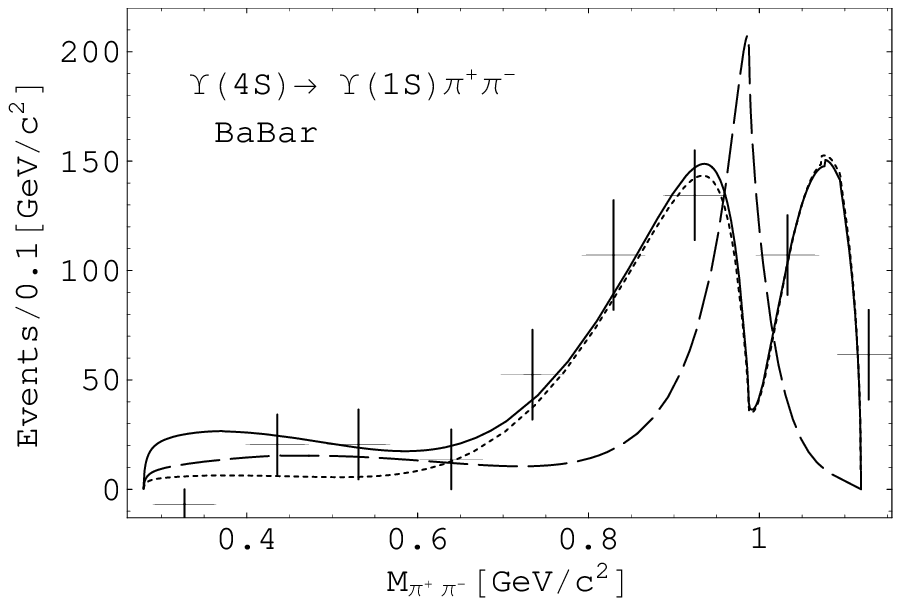}
\includegraphics[width=0.48\textwidth]{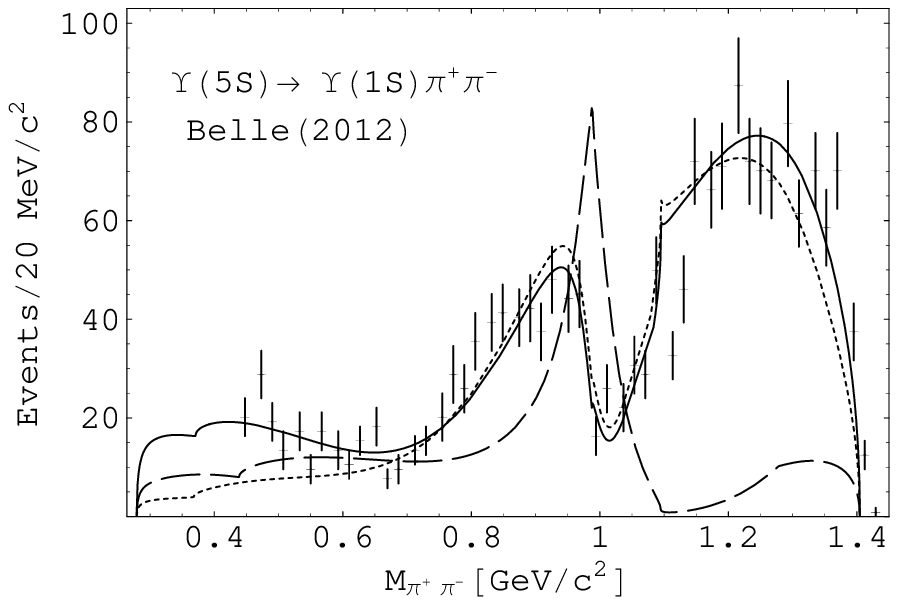}\\
\includegraphics[width=0.48\textwidth]{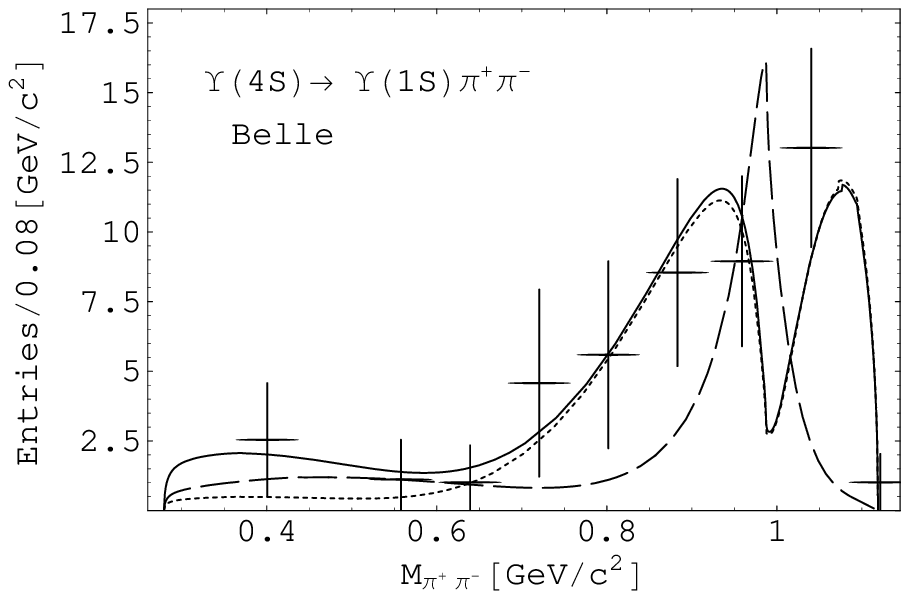}
\includegraphics[width=0.48\textwidth]{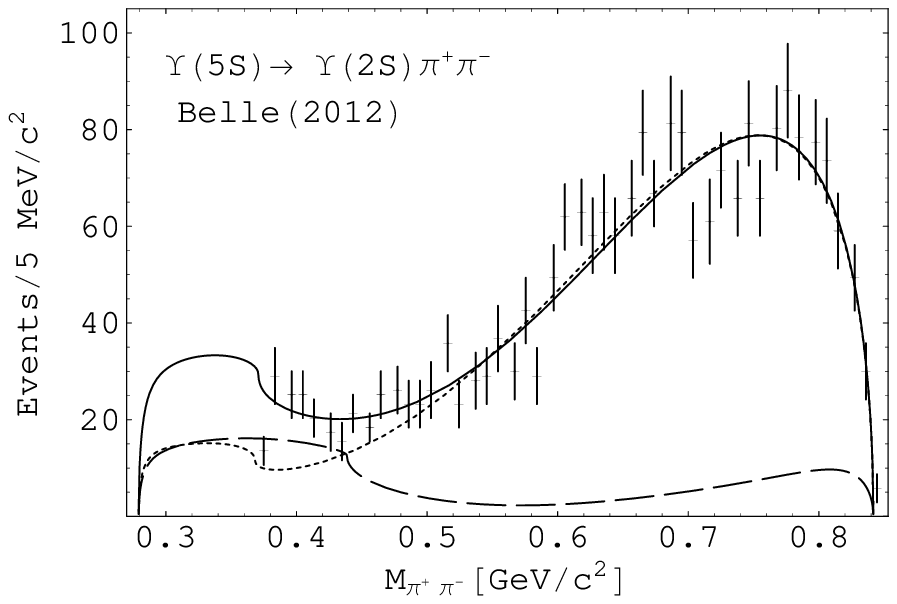}\\
\includegraphics[width=0.48\textwidth]{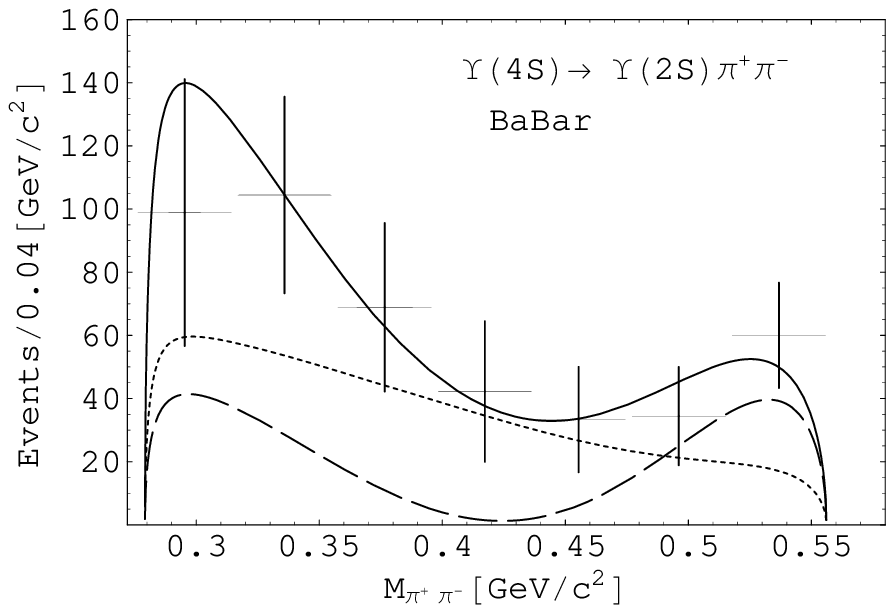}
\includegraphics[width=0.48\textwidth]{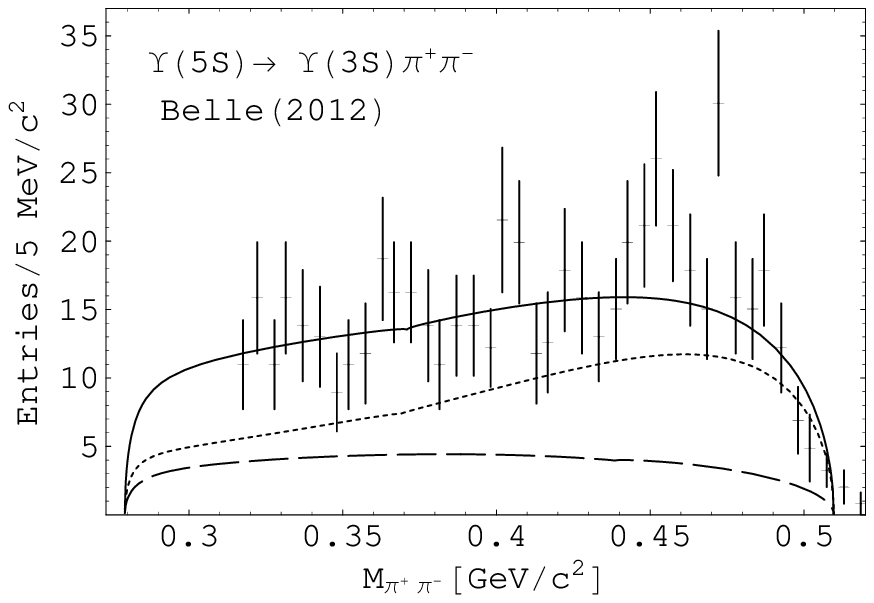}
\vspace*{-0.26cm}\caption{The decays $\Upsilon(4S)\to\Upsilon(1S,2S)\pi^+\pi^-$ (left-hand) and $\Upsilon(5S)\to\Upsilon(ns)\pi^+ \pi^-$ ($n=1,2,3$) (right-hand). The solid lines correspond to contribution of all relevant $f_0$-resonances;
the dotted, of the $f_0(500)$, $f_0(980)$, and $f_0^\prime(1500)$;
the dashed, of the $f_0(980)$ and $f_0^\prime(1500)$.}
\end{center}
\end{figure}

\section{Conclusions and Discussion}

The combined analysis was performed for the data on isoscalar S-wave processes
$\pi\pi\to\pi\pi,K\overline{K},\eta\eta$ and on the decays of the charmonia ---
$J/\psi\to\phi\pi\pi$, $\psi(2S)\to J/\psi\,\pi\pi$ --- and of the bottomonia ---
$\Upsilon(mS)\to\Upsilon(nS)\pi\pi$ ($m>n$, $m=2,3,4,5,$ $n=1,2,3$) from the ARGUS,
Crystal Ball, CLEO, CUSB, DM2, Mark~II, Mark~III, BES~II, {\it BaBar}, and Belle
Collaborations.

It is shown that the di-pion mass spectra in the above-indicated decays of charmonia
and bottomonia are explained by the unified mechanism which is based on our previous
conclusions on wide resonances \cite{SBLKN-jpgnpp14,SBLKN-prd14} and is related
to contributions of the $\pi\pi$, $K\overline{K}$ and $\eta\eta$ coupled channels
including their interference. It is shown that in the final states of these decays
(except $\pi\pi$ scattering) the contribution of coupled processes, e.g.,
$K\overline{K},\eta\eta\to\pi\pi$, is important even if these processes are energetically
forbidden.

Accounting for the effect of the $\eta\eta$ channel in the considered decays, both
kinematically (i.e. via the uniformizing variable) and also by adding the $\pi\pi\to\eta\eta$
amplitude in the formulas for the decays, permits us to eliminate nonphysical (i.e. those related
with no channel thresholds) non-regularities in some $\pi\pi$ distributions, which are present
without this extension of the description \cite{SBGKLN-pr15_2}. We obtained a reasonable
and satisfactory description of all considered $\pi\pi$ spectra in the two-pion transitions
of charmonium and bottomonium.

It was also very useful to consider the role of individual $f_0$ resonances in contributions
to the di-pion mass distributions in the indicated decays. For example, it is seen that the sharp
dips near 1~GeV in the $\Upsilon(4S,5S)\to\Upsilon(1S)\pi^+\pi^-$ decays are related with the
$f_0(500)$ contribution to the interfering amplitudes of $\pi\pi$ scattering and
$K\overline{K},\eta\eta\to\pi\pi$ processes.
Namely consideration of this role of the $f_0(500)$ allows us to make a conclusion on
existence of the sharp dip at about 1~GeV in the di-pion mass spectrum of
the $\Upsilon(4S)\to\Upsilon(1S)\pi^+\pi^-$ decay where, unlike
$\Upsilon(5S)\to\Upsilon(1S)\pi^+\pi^-$, the scarce data do not permit to draw such conclusions yet.

Also, a manifestation of the $f_0(1370)$ turned out to be interesting and unexpected.
First, in the satisfactory description of the $\pi\pi$ spectrum of decay $J/\psi\to\phi\pi\pi$,
the second large peak in the 1.4-GeV region can be naively explained as
the contribution of the $f_0(1370)$.
We have shown that this is not right -- the constructive interference between the contributions
of the $\eta\eta$ and $\pi\pi$ and $K\overline{K}$ channels plays the main role in formation of
the 1.4-GeV peak. This is quite in agreement with our earlier conclusion that the $f_0(1370)$
has a dominant $s{\bar s}$ component \cite{SBLKN-jpgnpp14}.

On the other hand, it turned out that the $f_0(1370)$
contributes considerably in the near-$\pi\pi$-threshold region of many di-pion mass distributions,
especially making the threshold bell-shaped form of the di-pion spectra in the decays
$\Upsilon(mS)\to\Upsilon(nS)\pi\pi$ ($m>n, m=3,4,5, n=1,2,3$).
This fact confirms, first, the existence of the $f_0(1370)$ (up to now there is no firm conviction
if it exists or not).
Second, that the exciting role of this meson in making the threshold bell-shaped form of the di-pion
spectra can be explained as follows: the $f_0(1370)$, being predominantly the $s{\bar s}$
state \cite{SBLKN-prd14} and practically not contributing to the $\pi\pi$-scattering amplitude,
influences noticeably the $K\overline{K}$ scattering;
e.g., it was shown that the $K\overline{K}$-scattering length is very sensitive to whether this
state does exist or not~\cite{SKN-epja02}.
The interference of contributions of the $\pi\pi$-scattering amplitude and the analytically-continued
$\pi\pi\to K\overline{K}$ and $\pi\pi\to\eta\eta$ amplitudes lead to the observed results.

It is important that we have performed a combined analysis of available data on the
processes $\pi\pi\to\pi\pi,K\overline{K},\eta\eta$, on decays of charmonia
$J/\psi\to\phi\pi\pi$, $\psi(2S)\to J/\psi(\pi\pi)$ and of bottomonia
$\Upsilon(mS)\to\Upsilon(nS)\pi\pi$ ($m>n, m=2,3,4,5, n=1,2,3$) from the ARGUS, Crystal Ball,
CLEO, CUSB, DM2, Mark~II, Mark~III, BES~II, {\it BABAR}, and Belle Collaborations.
The convincing description (including also the $\eta\eta$ channel) of practically all available
data on two-pion transitions of the $\Psi$ and the $\Upsilon$ mesons confirmed all our
previous conclusions on the unified mechanism of formation of the basic di-pion spectra,
which is based on our previous conclusions on wide resonances \cite{SBLKN-jpgnpp14,SBLKN-prd14}
and is related to contributions of the $\pi\pi$, $K\overline{K}$ and $\eta\eta$ coupled
channels including their interference. This also confirmed all our earlier results
on the scalar mesons~\cite{SBLKN-prd14}; the most important results are:

\begin{enumerate}

\item Confirmation of the $f_0(500)$ with a mass of about 700~MeV and
a width of 930~MeV (the pole position on sheet~II is
$514.5\pm12.4-465.6\pm5.9$~MeV).

\item
An indication that the $f_0(980)$ (the pole on sheet~II is
$1008.1\pm3.1-i(32.0\pm1.5)$~MeV) is neither a $q{\bar q}$ state nor the $K\overline{K}$ molecule.

\item
An indication for the ${f_0}(1370)$ and $f_0 (1710)$ to have a dominant
$s{\bar s}$ component.

\item
An indication for the existence of two states in the 1500-MeV region: the $f_0(1500)$
($m_{res}\approx1495$~MeV, $\Gamma_{tot}\approx124$~MeV) and
the $f_0^\prime(1500)$ ($m_{res}\approx1539$~MeV,
$\Gamma_{tot}\approx574$~MeV).

\end{enumerate}

This work was supported in part by the Heisenberg-Landau Program, by the Votruba-Blokhintsev Program for Cooperation of Czech Republic with JINR, by the Grant Agency of the Czech Republic (grant No. P203/15/04301), by the Grant Program of Plenipotentiary of Slovak Republic at JINR, by the Bogoliubov-Infeld Program for Cooperation of Poland with JINR, by the BMBF (Project 05P2015, BMBF-FSP 202), by Tomsk State University Competitiveness Improvement Program, the Russian Federation program ``Nauka'' (Contract No.\ 0.1526.2015, 3854), by Slovak Grant Agency VEGA under contract No.2/0197/14, and by the Polish National Science Center (NCN) grant DEC-2013/09/B/ST2/04382.


\end{document}